\input amstex
\documentstyle{amsppt}
\topmatter
\title The Appell hypergeometric functions \\ and classical separable mechanical
systems
\endtitle
\author Vladimir Dragovi\' c \endauthor
\rightheadtext{The Appell functions and separable systems}
\address Mathematical Institute SANU, Kneza Mihaila 35, 11000
Belgrade, Yugoslavia \endaddress
\email vladad\@ mi.sanu.ac.yu \endemail
\abstract
 A relationship between two old mathematical subjects is observed:
the theory of hypergeometric functions and the separability in classical mechanics.
Separable potential perturbations of the integrable billiard systems and the Jacobi
problem for geodesics on an ellipsoid are expressed through the Appell hypergeometric
functions $F_4$ of two variables. Even when the number of degrees of freedom
increases, if an ellipsoid is symmetric, the number of variables in the hypergeometric functions does not.
Wider classes of separable potentials are given by the obtained new formulae
automaticaly.
\endabstract
\endtopmatter
\footnote""{AMS Subject Classification (1991): Primary 58F05}
\document

\

\

\centerline{\bf 1. Introduction}

\

Appell introduced four families of hypergeometric functions of two variables in
1880's. Soon, he applied them in a solution of the Tisserand problem in the
celestial mechanics. The Appell functions have several other applications,
for example in the theory of algebraic equations, algebraic surfaces... The aim
of this paper is to point out the relationship between the Appell functions $F_4$
and another subject from classical mechanics - separability of variables in the
Hamilton - Jacobi equations.

The equation
$$
\lambda V_{xy}+3\left( yV_x-xV_y\right) +(y^2-x^2)V_{xy}+xy\left( V_{xx}-
V_{yy}\right)=0, \tag1
$$
appeared in Kozlov's paper [1] as a condition on function $V=V(x,y)$ to be
an integrable perturbation of certain type for billiard systems inside an
ellipse
$$
\frac {x^2}A + \frac {y^2}B = 1,\ \ \lambda=A-B.\tag2
$$
This equation is a special case of the Bertrand-Darboux equation [2,3,4]
$$
\aligned
(V_{yy}-V_{xx})(&-2axy-b'y-bx+c_1)+2V_{xy}(ay^2-ax^2+by-b'x+c-c')\\
&+V_x(6ay+3b)
+V_y(-6ax-3b')=0.
\endaligned\tag3
$$
It corresponds to the choise $a=-1/2,$ $b=b'=c_1=0,$ $ c-c'=-\lambda/2.$
The Bertrand-Darboux equation represents the necessary and suficient condition
for a natural mechanical system with two degrees of freedom
$$
H=\frac{1}2(p^2_x + p^2_y) + V(x,y)
$$
to be separable in the elliptical coordinates or some of their degenerations.

The solutions of the equation (1) in a form of the Laurent polynomials in
$x,y$ were described in [5]. The starting observation of this paper, that such
solutions are simply related to the well-known hypergeometric functions of the
Appell type is presented in the section 3. Such a relation automatically gives
a wider class of solutions of the equation (1). But what is more important,
it shows the existence of the connection between separability of the classical
systems on one hand, and the theory of hypergeometric functions on the other one.
The basic references for the Appell functions are [6,7,8].
Further, in section 3, the similar formulae for potential perturbations for
the Jacobi problem for geodesics on ellipsoid from [9] and for the billiard
systems on the surfaces with  constant curvature from [10], are given.

In the case of more than two degrees of freedom, the natural generalization
for the equation (1) is system (4). The system (4) for the three degrees of
freedom was derived in [11] in a connection with the billiard systems inside
an ellipsoid in $R^3$ and the Laurent polynomial solutions were given. In
section 4, we express these solutions for a symmeric ellipsoid through the hypergeometric functions.
Surprisingly, we obtain again the Appell hypergeometric functions of two
variables. The number of degrees of freedom increased but the number of
variables of the hypergeometric functions didn't!

The system 
$$
\aligned
&(a_i-a_r)^{-1}\left( x_i^2V_{rs}-x_ix_rV_{is}\right) =
(a_i-a_s)^{-1}\left( x_i^2V_{rs}-x_ix_sV_{ir}\right) \  i\ne r\ne s \ne i;\\
&(a_i-a_r)^{-1}x_ix_r \left( V_{ii}-V_{rr}\right) -\sum_{j\ne i,r}(a_i-a_j)^
{-1}x_ix_jV_{jr} \\
&+V_{ir}\left[ \sum_{j\ne i,r} (a_i-a_j)^{-1}x_j^2+(a_r-a_i)^{-1}(x_i^2-x_r^2)
\right] +V_{ir}\\
&+3(a_i-a_r)^{-1}\left( x_rV_i-x_iV_r\right)=0,  \ \ i\ne r,
\endaligned \tag4
$$
where $V_i=\partial V/\partial x_i$,
of $(n-1)\binom n2$   equations was formulated in [12] for arbitrary number of degrees of
freedom $n$. In [12] the generalization of the Bertrand-Darboux theorem
was proved. According to that theorem, the solutions of the system (4) are
potentials separable in generalized elliptic coordinates (see the Theorem
after lemma 2, below).

Some deeper explanation of the connection between the separability in the elliptic
coordinates and the Appell hypergeometric functions is not known yet.

\

\pagebreak

\centerline{\bf 2. Basic notations}

\

The function $F_4$ is one of the four hypergeometric functions in two
variables introduced by Appell [8,7] and defined as a serries:

$$
F_4(a,b,c,d;x,y)=\sum \frac{(a)_{m+n} (b)_{m+n}}{(c)_m (d)_n} \frac {x^m}{m!}
\frac{y^n}{n!},
$$
where $(a)_n$ is a standard Pochhammer symbol:

$$
\aligned
(a)_n&=\frac {\Gamma (a+n)}{\Gamma (a)}=a(a+1)\dots (a+n-1),\\
(a)_0&=1, \    
\endaligned
$$
(For example $ m!=(1)_m.$)

The series $F_4$ is covergent for $\sqrt x + \sqrt y \le 1.$ The functions $F_4$
can be continued to the solutions of the equations:
$$
\aligned
x(1-x)\frac {\partial ^2 F}{\partial x^2}- y^2 \frac {\partial ^2 F}{\partial
y^2}&-2xy \frac {\partial ^2F}{\partial x \partial y}+[c-(a+b+1)x]\frac {
\partial F}{\partial x} \\
&-(a+b+1)y\frac {\partial F}{\partial y}-ab F=0 ,
\endaligned 
$$
$$
\aligned
y(1-y)\frac {\partial ^2 F}{\partial y^2}- x^2 \frac {\partial ^2 F}{\partial
x^2}&-2xy \frac {\partial ^2F}{\partial x \partial y}+[c'-(a+b+1)y]\frac {
\partial F}{\partial y} \\
&-(a+b+1)x\frac {\partial F}{\partial x}-ab F=0 ,
\endaligned 
$$

\

\

\centerline{\bf 3. The separable systems with two degrees of freedom}

\

{\bf Billiard inside an ellipse}

\

\

Following [1,5] we will start with a billiard system which describes a particle
moving freely within an ellipse (2). At the boundary we assume elastic reflections
with equal impact and reflection angles. This system is completely integrable
and it has an additional integral
$$
K_1=\frac{\dot x^2}A+\frac{\dot y^2}B-\frac{(\dot x y -\dot y x)^2}{AB}.
$$
We are interested in a potential perturbations $V=V(x,y)$ such that the
perturbated system has an integral $\tilde K_1$ of the form
$$
 \tilde K_1=K_1 + k_1(x,y),
$$
where $k_1=k_1(x,y)$ depends only on coordinates. This specific condition
leeds to the equation (1) on $V$ (see [1]).

In [5] the Laurent polynomial solutions of the equation (1) were given.
The basic set of solutions consists of the functions
$$
V_k=\sum_{i=0}^{k-2} (-1)^i\sum_{s=1}^{k-i-1} U_{kis}(x,y,\lambda ,\alpha )+
\alpha y^{-2k},\ \ k\in N,
$$
$$
W_k=\sum_{i=0}^{k-2} \sum_{s=1}^{k-i-1}(-1)^s U_{kis}(y,x,\lambda ,\alpha )+
\alpha x^{-2k},\ \ k\in N,
$$
where
$$
U_{kis}=\binom {s+i-1}i \frac {[1-(k-i)][2-(k-i)]\dots [s-(k-i)]}{\lambda ^{s+i}
s!}\alpha x^{2s}y^{-2k+2i} .
$$
Now, we are going to rewrite the above formulae:
$$
\aligned
V_k&=\sum_{i=0}^{k-2} (-1)^i\sum_{s=1}^{k-i-1} U_{kis}(x,y,\lambda ,\alpha )+
\alpha y^{-2k},\ \ k\in N \\
&=\sum_{i=0}^{k-2} (-1)^i \sum_{s=1}^{k-i-1}\frac {\Gamma (s+i)\Gamma (s+i-k+1)}
{\Gamma (i+1)\Gamma (s)\Gamma (i-k+1)\Gamma (s+1)} x^{2s} y^{2(i-k)}+y^{-2k}\\
&=\frac 1{y^{2k}} \left( \sum_{i=0}^{k-2} \sum_{s=1}^{k-i-1} \frac {(1)_{s+i-1}
(2-k)_{s+i-1}}{i!(1)_{s-1} s! (1-k)_i}x^{2s} (-y^2)^i +1 \right) \\
&=\frac 1{y^{2k}} \left( x^2 \sum_{i=0}^{k-2}\sum_{s=0}^{k-i-2}
\frac {(1)_{s+i}(2-k)_{s+i}}{(2)_s (1-k)_i} \frac {(x^2)^s}{s!}\frac {(-y^2)^i}
{i!} +1 \right)\\
&=\frac 1{\tilde y^k}\left( \tilde x F_4(1; \ 2-k;\ 2, \ 1-k,\ \tilde x,\
-\tilde y) +1\right),
\endaligned
$$
where $\tilde x=x^2,\ \tilde y =-y^2$, and $F_4$ is the Appell function.
We just obtained the simple formula which expresses the potentials $V_k$,
for $k \in N$ from [5] through the Appell functions. We can use this formula
to spread the family of solutions of the equation (1) out of the set of
the Laurent polynomials.

Let $V(x,y)=\sum a_{nm}x^ny^m$. Then the equation (1) reduces to

$$
nma_{n,m}=(n+m)(ma_{n-2,m}-na_{n,m-2}).\tag5
$$

If one of the indices, for example the first one, belongs to $Z$, then $V$
doesn't have essential singularities. Put $a_{0,-2\gamma }=1$, where $\gamma $
is not necessary an integer.

If we define
$$
a_{\underbrace{2s+2}_{n},\underbrace{2i-2\gamma }_{m}}=\frac {(-1)^i (1)_{s+i}
(2-\gamma )_{s+i}}{(2)_s (1-\gamma )_i s! i!}.\tag6
$$

it can easily be seen that (6) is a solution of the eqation (5). So, let us denote 

$$
V_{\gamma }=\tilde y ^{-\gamma }\left( \tilde x F_4(1, 2-\gamma ,2 , 1-\gamma ,
\tilde x, \tilde y)+1\right). \tag7
$$
Then we have

\

{\bf Theorem 1} {\it Every function $V_{\gamma }$ given with (7) and $\gamma \in
C$ is a solution of the equation (1).}

\

Mechanical interpretation: with $\gamma \in R^-$ and the coefficient
multiplying $V_{\gamma }$ positive, we have potential barrier along 
$x$-axis; so we can consider a cut  along negative part of
$y$-axis.

\

{\bf The Jacobi problem for geodesics on an ellipsoid}

\

The Jacobi problem for the geodesics on an ellipsoid
$$
\frac{x^2}A + \frac{y^2}B + \frac{z^2}C=1
$$
has an additional integral
$$
K_1=(\frac{x^2}{A^2}+\frac{y^2}{B^2}+\frac{z^2}{C^2}) (\frac{\dot x^2}A+\frac{\dot y^2}B+\frac{\dot z^2}C).
$$
The potential perturbations $V=V(x,y,z)$ such that perturbated systems have the
integrals of the form
$$
\tilde K_1 = K_1 + k(x,y,z)
$$
satisfy the following system (see [9])
$$
\aligned
&\left( \frac {x^2}{A^2}+\frac {y^2}{B^2}+\frac {z^2}{C^2}\right) V_{xy} 
\frac {A-B}{AB}-3\frac y{B^2}\frac {V_x}A+3\frac x{A^2}\frac {V_y}B +\left( 
\frac {x^2}{A^3}-\frac {y^2}{B^3}\right) V_{xy}+\\
&+\frac {xy}{AB}\left( \frac {V_{yy}}A-\frac {V_{xx}}B\right) +\frac {zx}{CA^2}
V_{zy}-\frac {zy}{CB^2}V_{zx}=0 
\endaligned
$$

$$
\aligned
&\left( \frac {x^2}{A^2}+\frac {y^2}{B^2}+\frac {z^2}{C^2}\right) V_{yz} 
\frac {B-C}{BC}-3\frac z{C^2}\frac {V_y}B+3\frac y{B^2}\frac {V_z}C +\left( 
\frac {y^2}{B^3}-\frac {z^2}{C^3}\right) V_{yz}+\\
&+\frac {yz}{BC}\left( \frac {V_{zz}}B-\frac {V_{yy}}C\right) +\frac {xy}{AB^2}
V_{xz}-\frac {xz}{AC^2}V_{xy}=0 
\endaligned \tag8
$$

$$
\aligned
&\left( \frac {x^2}{A^2}+\frac {y^2}{B^2}+\frac {z^2}{C^2}\right) V_{zx} 
\frac {C-A}{AC}-3\frac x{A^2}\frac {V_z}C+3\frac z{C^2}\frac {V_x}A +\left( 
\frac {z^2}{C^3}-\frac {x^2}{A^3}\right) V_{zx}+\\
&+\frac {xz}{AC}\left( \frac {V_{xx}}C-\frac {V_{zz}}A\right) +\frac {zy}{BC^2}
V_{xy}-\frac {yx}{BA^2}V_{yz}=0 
\endaligned 
$$
The system (8) replaces in this problem the equation (1). The solutions
of the system in the Laurent polynomial form were found in [9]. We can
transform them in a following way.

$$
\aligned
&V_{l_0}(x,y,z)=\sum_{0\le k\le s, k+c\le l_0} (-1)^s \binom {s+k-1}k
(x^2)^{-l_0+k}(y^2)^s(z^2)^{l_0-(k+s)-1}\\
&\times \frac {c^{s+k}(c-a)^s(c-b)^k 2^{k+s}(-l_0+1)\dots (-l_0+(k+s))}
{b^ka^s (b-a)^{k+s}2^s2^k s! (-l_0+1)\dots (-l_0+k)} 
(z^2)^{l_0-(k+s)-1}\\
&=\sum \frac{(s+k-1)!(-l_0+1)(-l_0+2)_{s+k-1}(z^2)^{l_0}}
{k!(s-1)!s! (-l_0+1)_k (x^2)^{l_0}}
\left[ \frac {x^2c(a-c)}{z^2(b-a)a}\right] ^s \left[ \frac {y^2c(c-b)}{z^2
(b-a)b}\right]^k\\
&=(-l_0+1)\left( \frac {z^2}{x^2}\right)^{l_0} \sum \frac {(1)_{s+k-1}(-l_0+2)
_{s+k-1}}{(2)_{s-1}(-l_0+1)_k}\hat x^s \hat y^k\\
&=(-l_0+1)\left( \frac {z^2}{x^2}\right)^{l_0}F_4(1;-l_0+2;2,-l_0+1,\hat x,
\hat y),
\endaligned
$$
where
$$
\frac {x^2c(a-c)}{z^2(b-a)a}= \hat x, \ \  \frac {y^2c(c-b)}{z^2(b-a)b}
= \hat y
$$
In the above formulae $l_0$ was an integer. We have strightforward generalization:

\

{\bf Theorem 2.} {\it For every $\gamma \in C$ the function
$$
V_{\gamma}=(-\gamma +1)\left( \frac {z^2}{x^2}\right)^{\gamma}F_4(1;-\gamma +2;2,-\gamma +1,\hat x,
\hat y),
$$
is a solution of the system (8).}

\

\

{\bf Billiard systems on the constant curvature surfaces}

\

Potential perturbations of the billiard systems on the constant curvature
surfaces were analysed in [10]. Following the notation from [10], let the
billiard $D_S$ be a subset of the surface $\Sigma _S,$ of the curvature
$S=+1$ or $S=-1,$ bounded with the quadric
$Q_S$, where
$$
\aligned
\Sigma_+&=\{ r=(x,y,z)\in R^3 \ | \ \langle r,r\rangle _+=1\}, \
\langle r_1, r_2\rangle _+=x_1y_1+x_2y_2+x_3y_3;\\
\Sigma_-&=\{ r=(x,y,z)\in R^3 \ | \ \langle r,r\rangle _-=-1, \ z>0\}, \
\langle r_1, r_2\rangle _-=x_1y_1+x_2y_2-x_3y_3;\\
Q_S&=\Sigma_S \cap \{ r\in R^3\ | \ \langle Qr,r\rangle _s=0\} \ne \emptyset, \
Q=\text{diag} \left( \frac 1A , \frac 1B ,\frac 1C \right).
\endaligned
$$
Then the billiard system has the integral
$$
K= \frac{(\dot x y -\dot y x)^2}{AB} + S\frac{(\dot x z -\dot z x)^2}{AC} + S\frac{(\dot z y -\dot y z)^2}{BC}
$$
As before we are looking for  potentials $V=V(x,y,z)$ such that the pertubated
system has an integral of the form
$$
\tilde K = K + k(x,y,z).
$$
In this case the condition is given by the system [10]:
$$
\aligned
3CyV_x&-3CxV_y+V_{xy}(C(y^2-x^2)+Kz^2(B-A))\\
&+cxyV_{xx}-cxyV_{yy}+azyV_{zx}-
bzxV_{zy}=0,
\endaligned
$$
$$
\aligned
3BzV_x&-K3BxV_z+V_{xz}(B(z^2-Kx^2)+Ky^2(C-A))\\
&+BzxV_{xx}-KBzxV_{zz}+AzyV_{xy}-
KCyxV_{yz}=0,
\endaligned \tag 10
$$
$$
\aligned
3AzV_y&-K3AyV_z+V_{yz}(A(z^2-Ky^2)+Kx^2(C-B))\\
&+AzyV_{yy}-KAzyV_{zz}+BzxV_{xy}-
KCxyV_{xz}=0.
\endaligned
$$
Starting from the solutions from [10]
$$
V_{l_0}=\frac 1{z^{2l_0}}\sum \Sb 0\le k\le l_0-1\\ 0\le m\le l_0-k-1\endSb
a_{m,k} x^{2m} y^{2l_0-2-2k-2m} z^{2k},
$$
where
$$
a_{m,k}=K^{l_0-k-1}\left( \frac {C-B}{C-A}\right) ^m \binom {l_0-k-1}m \binom {k+m-1}k
\left( \frac {A-B}{C-A}\right) ^k ,
$$
we come to

\

{\bf Theorem 3.} {\it The functions
$$
V_{\gamma}={\hat y}^{- \gamma}(x^2 F_4(1,2-\gamma,2,1-\gamma,\hat x,\hat y)+1),
$$
where
$$
\frac {x^2(B-C)}{y^2(C-A)}= \hat x, \ \  K\frac {z^2(A-B)}{y^2(C-A)}
= \hat y
$$
are solutions of the system (10), for $\gamma \in C.$}

\

\

\centerline {\bf 4. More than two degrees of freedom}

\

In the previous section we saw that the integrable perturbations of
separable systems with two degrees of freedom led to the hypergeometric
functions of two variables. Now, one can expect that in a case of more
than two degrees of freedom, the integrable potentials are connected with the 
hypergeometric functions again, but with more than two variables. We
will consider the billiard system inside an ellipsoid in $R^3$, and we
will see that corresponding potential perturbations are still related
to the Appell function $F_4$ of two variables, if the ellipsoid is symmetric.

\

{\bf Billiards inside an symmetric ellipsoid in $R^3$}

\

Let us consider the billiard system within an ellipsoid in $R^3$
$$
\frac {x^2}A + \frac {y^2}B + \frac {z^2}C = 1.
$$
Potential perturbations $W=W(x,y,z)$ of such systems in a form of
the Laurent polynomials were calculated in [11]. They satisfy the
system (4) for $n=3$.

$$
W_{l_0}=\frac 1{z^{2l_0}}\sum_{0\le m+n+k<l_0}\frac {(l_0-k-1)!(-1)^n}{m!n!
(l_0-1-k-m-n)!}\frac {P^k_{m,n}(\beta , \gamma )}{\gamma ^{m+k}\beta ^{n+k}}
x^{2m}y^{2n}z^{2k},\tag11
$$
where
$$
P_{m,n}^k(\beta ,\gamma )=\sum_{i=0}^k \binom {m+k-1-i}{k-i}\cdot \binom
{n+i-1}i (-1)^i \beta ^{k-i} \gamma ^i.
$$
and $\beta = B-C, \ \gamma = C-A.$

The symmetric case $A=B$ corresponds to the condition $\gamma + \beta = 0.$

\

{\bf Lemma 1.} {\it If $\gamma +\beta =0$ then we have}
$$
P_{m,n}^k(\beta ,-\beta )=\binom {k+m+n-1}k \beta ^k .\tag12
$$

\
{\bf Proof.}
$$
\aligned
P_{m,n}^k (\beta ,-\beta )&=\beta ^k \sum_{i=0}^k  \binom {m+k-i-1}{k-i}
\binom {n+i-1}i \\
&=\beta ^k \sum_{i=0}^k \binom {k-i+m-1}{k-i}\binom {i+n-1}i =\binom {k+m+n-1}k
\beta ^k .
\endaligned
$$

\

By putting (12) into (11) we get, using $\gamma =-\beta $
$$
\aligned
&W_{l_0}=C\frac 1{\hat z^{l_0}}\sum_{m+n+k=1}^{l_0}
\frac{(l_0-1)\dots (l_0-m-n-k)(m+n+k-1)!(-1)^n}{(l_0-1)\dots (l_0-k) (m+n-1)!}
\frac {\hat x^m}{m!}\frac {\hat y
^n}{n!} \frac {\hat z^k}{k!}\\
&=\frac {C}{(\hat z)^{l_0}}\left[ \sum_{m+n+k=1}^{l_0}\frac {(-l_0+2)\dots
(-l_0+m+n+k) (1)_{m+n+k-1}}{(-l_0+1)\dots (-l_0+k)
(-1)^{-m} (1)_{m+n-1}} \frac {\hat x^m}{m!}\frac {\hat y^n}{n!}\frac {\hat z^k}
{k!}+1\right]
\endaligned 
$$
where $\hat x=\frac {x^2}{\gamma },\ \hat y =\frac {y^2}{\beta }, \
\hat z =\frac {z^2}{\gamma }$.

$$
\aligned
&W_{l_0}=C\cdot \hat z ^{-l_0}\left( \sum_{m+n+k=1}^{l_0}\frac {(1)_{m+n+k-1}
(2-l_0)_{m+n+k-1}(-1)^m}{(1)_{m+n-1}(1-l_0)_k} \frac {\hat x^m}{m!}\frac {\hat y^n}{n!}\frac {\hat z^k}
{k!}+1\right) \\
&= C\cdot \hat z ^{-l_0}\left( \sum_{m+n+k=1}^{l_0}\frac {(1)_{m+n+k-1}
(2-l_0)_{m+n+k-1}(-1)^m (m+n)!}{(2)_{m+n-1}(1-l_0)_k (m+n-1)!} \frac {\hat x^m}{m!}\frac {\hat y^n}{n!}\frac {\hat z^k}
{k!}+1\right) \\
&=C\cdot \hat z ^{-l_0}\left( \sum_{m+n+k=1}^{l_0}\frac {(1)_{m+n+k-1}
(2-l_0)_{m+n+k-1}}{(2)_{m+n-1}(1-l_0)_k} \frac {(-\hat x+\hat y)^{m+n}}{(m+n-1)!}
\frac {\hat z ^k}{k!}+1\right)\\
&=C\cdot \hat z ^{-l_0}\left[ (-\hat x +\hat y)\sum_{0\le s+k<l_0}\frac
{(1)_{s+k}(2-l_0)_{s+k}}{(2)_s(1-l_0)_k} \frac {(-\hat x+\hat y)^s}{s!}
\frac {\hat z ^k}{k!}+1\right],
\endaligned
$$
where $C=(1-l_0)\gamma ^{-l_0}$, $m+n-1=s$.

\

{\bf Theorem 4.} {\it Generalizations of the integrable potential perturbations
from [11] in the symmetric case are given by:}
$$
W_{l_0}=(\hat z)^{-l_0} \left[ (-\hat x +\hat y)F_4(1,2-l_0;2,1-l_0; -\hat x
+\hat y, \hat z)+1\right],
$$
where $l_0\in C$.

\

\

{\bf The case of general dimension}

\

We consider billiard system in $R^n$ within an ellipsoid
$$
\frac{x_1^2}{a_1} +\dots + \frac{x_n^2}{a_n} =1.
$$
For $n\ge 3$ we start from a separable system with $n$ integrals $K_1,...,K_n$
which are mutualy in involution, where $K_n=H$ is a Hamiltonian and
$$
K_i=\sum_{j\ne i} \frac{(\dot x_i x_j -\dot x_j x_i)^2}{a_i-a_j},\ \ i=1,...,n-1.\tag13
$$
Then we are interested
in potential perturbations $k_1,...,k_n$, where $k_i=k_i(x_1,...,x_n)$ depend
only on coordinates and $V=k_n$. The conditions
$$
\{\tilde K_n, \tilde K_i\}=0, \ \ i=1,...,n-1,\tag14
$$
where
$$
\tilde K_i=K_i+k_i, \ \ i=1,...,n,\tag15
$$
are equivalent to the system (4).
Nontrivial (and maybe unexpected) fact is that new integrals commute between
themselves.

\

{\bf Lemma 2} {\it From (13) and (14), it follows that}
$$
\{\tilde K_j, \tilde K_i\}=0, \ \ i,j=1,...,n-1.
$$

\

This was checked by direct calculation for $n=3$ in [11], but in general case
it follows from the generalized Bertrand- Darboux theorem proved in [12]:

\

{\bf Generalized Bertrand-Darboux Theorem} [12] {\it For a natural Hamiltonian 
system with a Hamiltonian
$$
H=\frac{1}2\sum_{i=1}^n p_i^2 + V(x)
$$
the following three conditions are equivalent:

\item {(a)} It has $n-1$ global, indipendent, involutive integrals of the
form (15).

\item {(b)} The potential $V$ satisfies the system (4).

\item {(c)} The Hamilton - Jacobi equation for $H$ is separable in generalized
elliptic coordinates $(u_1,...,u_n)$ given by
$$
1+\sum_{i=1}^n\frac{x_i}{z-a_i}=\frac {\prod_{j=1}^n (z-u_j)}{\prod_{k=1}^n
(z-\alpha _k)}.
$$

\

\

{\bf Theorem 5} {\it Integrable potential perturbations of the billiard
systems inside a symetric ellipsoid in $R^n$ for arbitrary $n \in N$ are given
with}
$$
V_k^n=(\hat x_n)^{-k}\left[ (\hat x_1+\dots +\hat x_{n-1})F_4(1,2-k;2,1-k;
 \hat x_1+\dots +\hat x_{n-1}, \hat x_n)+1\right]
 $$
for any $k\in C$.}

\

\

\centerline {\bf 5. Conclusion}

\

If we denote in (11) $-\beta / \gamma $ as $q$, then, theorem 4 shows
that the potentials $W_{l_0}$ are certain deformations of the (Appell)
hypergeometric functions. The analysis of this  deformation and comparison
to the known $q$-deformations and generaisations of the hypergeometric functions
multivariable ([6]) remains as an interesting problem.

Potential perturbations of the classical nonholonomic rigid body problems are
described in [13]. It looks that they are also connected with the  hypergeometric
functions.

From geometric point of view, it is well known that the billiard systems within
an ellipse are closely related to the Poncelet and the Cayley theorem [14-16].
So, the Appell hypergeometric functions define natural deformations of these
classical projective geometry settings.

\

\

{\bf Acknowledgment} This research was partialy done during the author's
visit to the Mathematical Department of the Kyoto University as a Matsumae
International Foundation Fellow. It is a great pleasure to thank professor
M. Jimbo for warm hospitality and A.Yu. Orlov for valuable discussions.

\

\

\centerline {\bf References}

\

\item {[1]} Kozlov V.V.: {\it Some integrable generalizations of the
Jacobi problem for the geodesics on the ellipsoid} (in Russian), Prikl.
Mat. Mekh., Vol. {\bf 59}, No.1, 3-9, (1995.)

\item {[2]} Whittaker E.T.: {\it A treatise on the analitical dynamics
of particles and rigid bodies, thierd edition,} Cambridge, the University
press (1927)

\item {[3]} Bertrand J., Journ.de Math., Vol. 17, p. 121, (1852)

\item {[4]} Darboux G.: Archives Neerlandaises (2), Vol. 6, p. 371 (1901)

\item {[5]} Dragovic V.: {\it On integrable potentials of billiard within
ellipse} (in Russian), Prikl. Mat. Mekh., Vol. {\bf 62}, (1998.), No. 1, 166-169.

\item {[6]} Vilenkin N. Ja. and Klimyk A. U.:{\it Representation of Lie groups and
special functions}, Recent Advances, Kluwer Academic Rublishers, p. 497 (1995)

\item {[7]} Appell P. and Kampe de Feriet J.: {\it Fonc\-tions hyper\-geo\-met\-riques
et hy\-per\-sphe\-ri\-ques.} Polynomes d'Hermite, Gauthier Villars, Paris (1926)

\item {[8]} Appell P. {\it Sur les fonctions hypergeometriques de deux variables
et sur des equations lineaires aux derivees partielles}, Comptes Rendus, v.90,
p. 296 (1880)

\item {[9]} Dragovic V.: {\it On integrable perturbations of the Jacobi
problem for the geodesics on the ellipsoid}, J. of Phys. A, Math. and General,
Vol. {\bf 29}, L317-L321, (1996.)

\item {[10]} Jovanovic B.: {\it Integrable perturbation of billiards on
constant curvature surfaces}, Phys. Lett. A, Vol. {\bf 231}, 353-358, (1997.)

\item {[11]} Dragovic V., Jovanovic B.: {\it On integrable potential
perturbations of billiard system within ellipsoid}, J. of Math. Phys.,
Vol. {\bf 38}, (1997.)

\item {[12]} Marshall I., Wojciechowski S.:{\it  When is a Hamiltonian system Separable?}

\item {[13]} Dragovi\' c V., Gaji\' c B., Jovanovi\' c B.:{\it  Generalizations of
classical integrable nonholonomic rigid bodu sustems}, J. Phys A: Math. Gen.,
Vol. 31, 9861-9869 (1998)

\item {[14]} H. Lebesgue {\it Les coniques }, Gauthier-Villars, Paris, (1942),
115-149.

\item {[15]} P. Griffiths, J. Harris {\it On Cayley's explicit solution to 
Poncelet's porism}, L'En\-se\-ignement mathem. {\bf 24}, (1978),  31-40.

\item {[16]} Dragovi\' c V., Radnovi\' c M. {\it On periodical trajectories
of the billiard systems within an ellipsoid in $R^d$ and generalized Cayley's 
condition}  J. of Math. Phys.,Vol. {\bf 39}, No. 11, (1998)

\enddocument